# Strong surround antagonism in the dLGN of the awake rat

Balaji Sriram[1] and Pamela Reinagel[1*]

1. Division of Biology, University of California, San Diego, California

**Corresponding Author**: Pamela Reinagel, 9500 Gilman Drive MC#0357, La Jolla CA 92093.

**Email**: preinagel@ucsd.edu

**Draft Date:** April 16, 2012

# Abstract

Classical center-surround antagonism in the early visual system is thought to serve important functions such as enhancing edge detection and increasing sparseness. The relative strength of the center and surround determine the specific computation achieved. For example, weak surrounds achieve low-pass spatial frequency filtering and are optimal for denoising when signal-to-noise ratio (SNR) is low. Balanced surrounds achieve band-pass spatial frequency filtering and are optimal for decorrelation of responses when SNR is high. Surround strength has been measured in the retina and dorsal Lateral Geniculate Nucleus (dLGN), primarily in anesthetized or ex vivo preparations. Here we revisit the center-surround architecture of dLGN neurons in the un-anesthetized rat. We report the spatial frequency tuning responses of N=47 neurons. We fit these tuning curves to a difference-of-Gaussians (DOG) model of the spatial receptive field. We find that some dLGN neurons in the awake rat (N=8/47) have weak surrounds. The majority of cells in our sample (N=29/47), however, have well-balanced center and surround strengths and band-pass tuning curves. We also observed several neurons (N=10/47) with notched or dual-band-pass tuning curves, a response class that has not been described previously. Within the space of circularly concentric DOG models, strong surrounds were necessary and sufficient to explain the dual-band-pass spatial frequency tuning of these cells. It remains to be determined what advantage if any is conferred by this novel response class, or by the heterogeneity of surround strength as such. We conclude that surround antagonism can be strong in the dLGN of the awake rat.

# Introduction

One of the most striking and consistent feature of early stages of sensory processing is the presence of center-surround antagonism. First described in the retina (Kuffler 1953) and dLGN (Hubel and Wiesel 1961), this manifests as neurons having concentric receptive fields with a center (classified as ON or OFF depending on whether the neuron increases its firing rate to an increase or decrease in luminance at the center) and an antagonistic surround (which is larger than the center and has luminance preference opposite to that of the center). Center-surround antagonism is also found in other sensory modalities such as in the auditory periphery (Knudsen and Konishi 1978), in the somatosensory cortex (DiCarlo et al. 1998), and in the whisker barrel system(Bellavance et al. 2010; Simons and Carvell 1989).

Early theories about the function of center-surround antagonism include edge enhancement (Hartline et al. 1956) and redundancy reduction (Attneave 1954; Barlow 1961). Testing and extending these theories remains an active experimental and theoretical field (Atick 2011; Atick JJ 1990; Dan et al. 1996; Graham et al. 2006; Kuang et al. 2012; Olshausen and Field 1996; Pitkow and Meister 2012; Puchalla et al. 2005; van Hateren 1992)

The computational effect of surround antagonism depends on the relative strength and size of the center and surround. For example, weak surrounds achieve low-pass spatial frequency filtering and are optimal for de-noising when signal-to-noise ratio (SNR) is low, while balanced surrounds achieve band-pass spatial frequency filtering and are optimal for de-correlation of responses when SNR is high (Atick JJ 1990). The strength and size of both the center and surround of the receptive fields of neurons in the retina and the dLGN has been measured in multiple species (Alitto et al. 2011; Cheng et al. 1995; Dacey et al. 2000; Grubb and Thompson 2003; Heine and Passaglia 2011; O'Keefe et al. 1998; Xu et al. 2002). These studies all find a distribution of surround strengths ranging from weak to balanced, with an average surround about 75% the strength of the center.

Rodents are becoming an important model to study visual circuits due to their availability and relative inexpensiveness, their ability to perform complex behaviors (Busse et al. 2011; Creer et al. 2010; Harvey et al. 2012; Meier et al. 2011; Zoccolan et al. 2009), their applicability to studying visual diseases (Sekirnjak et al. 2011) and the relative ease of applying genetic techniques (Morozov 2008; Thomas and Capecchi 1987) to modify circuit function. The functional properties of the early visual system of rodents have been characterized since the 1960s (Anderson et al. 1977; Fukuda et al. 1979; Kriebel 1975; Lennie and Perry 1981), but the strength of surround antagonism has not been studied in detail.

## Experimental Procedures

All procedures were conducted with the approval and under the supervision of the Institutional Animal Care and Use Committee at the University of California San Diego. Six male Long-Evans rats (Harlan) were used for this study. A preliminary account of these physiology methods, including surgical implant design, head fixed recording methods, integration with stimulus display and eye-tracking hardware and software, has been presented previously (Flister and Reinagel 2010; Sriram et al. 2011)

**Surgery**: Adults hooded male rats (*Rattus norvegicus*, >P90) are deeply anesthetized using 5% isoflurane. Ringers solution (15 ml/kg ) is provided for hydration and Atropine (0.05 mg/kg) is injected to control secretion. The scalp is shaved and sterilized with 70% isopropyl alcohol /Betadine pads. The rat is placed on a stereotaxic apparatus and Sensorcaine (0.1 ml) injected into the scalp. An incision is made over the left dLGN (4.5mm Posterior, 3.5mm Lateral from Bregma). After removing the fascia over the skull, a craniotomy is drilled over the putative dLGN center (dimensions: 1.5-2 mm M-L, 2-3 mm A-P). An eppendorf tube is glued over the craniotomy to provide easy access for future recording. Titanium screws (5 or 6) are placed over the exposed skull and the craniotomy filled with cement. A hex standoff and an aluminum spacer are also included in the head cap for future head fixing.

**Head-fixing rats**: Male rats are constrained in a sock and injected with a mild sedative, Midazolam(0.3-0.6 ml/kg), 10 minutes prior to being placed on the rig. Rat's heads are fixed with screws threaded through the standoff and the spacer. In our hands, this led to stable recording sessions lasting about 120 minutes and allowed for stable recording of single units lasting 10-20 minutes in duration. Each rat can be used for recording for a duration between 3-

12 weeks.

**Stimulus presentation**: One of two monitors was used to present visual stimuli to the right eye while recording from putative single units in the contralateral dLGN. Both monitors (Westinghouse L2410NM(LCD)/Sony Trinitron PF790-VCDTS21611(CRT)) refreshed at 60 Hz and were linearized multiple times throughout the experiment's duration. Results were similar in the recordings collected with both monitors. Custom Matlab (Mathworks Inc., Waltham, MA) scripts were used to present either stochastic noise stimuli or drifting gratings stimuli to the right eye of the rat (Meier et al. 2011). Our stimuli were constructed using the Psychophysics toolbox (Brainard 1997; Kleiner M 2007; Pelli 1997).

**Electrophysiological recording**: The semi-chronic well was exposed, and an extracellular electrode (Tungsten, FHC, Bowdoin, ME or pulled Quartz microelectrodes filled with ringers solution) was inserted into stereotaxically defined coordinates (4.5P,3.5L,5-6V) in the rat brain. Voltage traces were amplified (10-1000X) and filtered (1Hz-10kHz) (AM1800; AM Systems, Sequim, WA), digitized (NIDAQ PCI-6259, National Instruments, Austin, TX), and stored in a local computer. After each session, the semi-chronic well was washed well with antibiotic solution (Baytril 0.05mg/ml), cleaned well with neutral saline and plugged with silicone gel until the next session.

**Single Unit Isolation:** Single units were identified as large negative (tungsten,FHC) or positive (quartz microelectrode) deflecting spikes with little visible in terms of back ground activity. Rough sorting criteria were used to characterize the single unit on-the-fly. All analyses shown were performed on more stringent offline sorting using Klustakwik (Harris et al. 2000). Units were kept for analysis if they met several criteria: (1) thresholded spike waveforms were aligned at the positive peak and all waveforms formed a well isolated cluster; (2) spike shape was relatively constant spike shape; and (3) no refractory violations. Spike amplitude variation was common for well-isolated single units, especially during bursts-like epochs. Care was taken to include as many of these burst-like spikes as possible and no distinction was made between burst-like spikes and other spikes.

**Eye stability and quality**: The rat's eye position was monitored using an infrared tracker (Eyelink 1000, SR Research, Kanata, Ontario, Canada). Tracking data was digitized using the EyeLink Toolbox (Cornelissen et al. 2002). When aroused, rats perform low-amplitude (<5°) saccades (Hikosaka and Sakamoto 1987) (but see (Chelazzi et al. 1989)). In our preparation, these saccades are infrequent (< 1 Hz) when aroused and even less frequent when dormant. After saccades, the eye position typically decayed back to the central fixation point. Preliminary experiments showed that rats maintain fixation within a 5° circle around the mean eye position > 65% of the time (data not shown). This was sufficient for our purposes, which require only that the receptive field fall within a full-field grating stimulus. We used spike-triggered-averages to locate the center of the receptive field at the mean eye position (see **Single Unit Characterization**). Clear artificial tears were applied to the eye to keep them moist. Excess tears were removed using a cotton swab.

**Single Unit Characterization**: Once single units were identified as visually driven (using an opthalamoscope) the position of the rat relative to the monitor was adjusted such that the receptive field of the unit was within the extent of the monitor. White noise stimuli were used to locate the center of the receptive field within the monitor. Once the location of the receptive field was identified, vertical drifting gratings of varying spatial frequencies was presented to the rat on a linearized CRT monitor. Gratings were at full contrast and drift frequency was 2 or 4 Hz with a refresh rate of 60 frames/s. This drift frequency was chosen so as to drive maximal neural responses without engaging the Optokinetic Reflex (Fuller 1985). Each "trial" consisted of a 2 or 3 s presentation of a constant spatial frequency; spatial frequencies were interleaved in pseudorandom order without gaps between until each stimulus was presented once, and this sequence was repeated three or more times.

**Spatial Frequency Tuning**: The responses of each single unit were temporally discretized at the stimulus refresh rate (60 Hz). To mitigate the effects of small eye movements on stimulus phase, response power estimates were calculated on a trial by trial basis as the Fourier transform of the autocorrelation function (Wiener-Khinchin theorem). This was repeated for each trial for a specific spatial frequency and across all spatial frequencies. The f1 response amplitude for each trial was measured as the square root of the power at the stimulus temporal frequency. Care was taken to ensure that the measured amplitude actually corresponded to a peak in the spike train power spectrum. If no peaks were visible, the single unit was rejected from the analysis. Non-stationary units, showing inconsistent tuning curves or large changes in mean rate between repeats, were rejected from the analysis.

**Fitting DOG Model:** Custom MatLab routines were used to fit f1 responses with a modified difference of Gaussians (DOG) model (Enroth-Cugell and Robson 1966; Grubb and Thompson 2003):

$$R_{f_s} = min(R_{f_s}) + \left| K_c \pi r_c^2 e^{-\pi^2 r_c^2 f_s^2} - K_s \pi r_s^2 e^{-\pi^2 r_s^2 f_s^2} \right| \quad r_c < r_s; K_c > K_s \qquad \textit{Equation 1}$$

The base offset (removing residual power to get the minimum f1 value to 0) improved the quality of fits. The absolute value function is used because the measured power is constrained to be positive. The Simplex Search Algorithm (`fmincon` in Matlab) was used to search for the correct combination of parameters ($K_c$, $K_s$, $r_c$, $r_s$) that best fit the response under a constrained optimization protocol. For each single unit, fitting was done to three separate non-linear constraints, $\eta < 1, \eta = 1$ and $\eta > 1$ where,

$$\eta = \frac{K_s r_s^2}{K_c r_c^2} \qquad \text{Equation 2}$$

is the ratio of the integrated weight in the surround to the integrated weight in the center (Xu et al. 2002); (Croner and Kaplan 1995); (Enroth-Cugell and Robson 1966); (Grubb and Thompson 2003). This provided three separate classes of solutions to which each single unit belonged. The fitting algorithm minimized the sum of the squares of the difference between fitted values and the actual values (Figure 1).

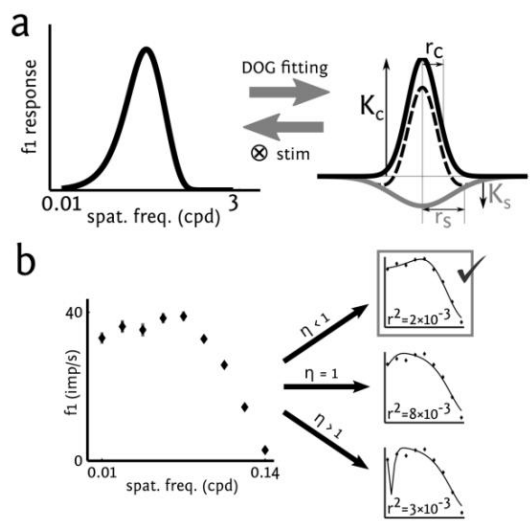

**Figure 1. Difference of Gaussian Model Fitting.** (a) A model receptive field is defined by two circular, concentric 2D Gaussian densities, a cross section through which is shown at right. The receptive field center (solid black curve) is defined by the radius $r_c$ and peak amplitude $K_c$ of the smaller Gaussian, the sign of which determines the response type (ON or OFF) of the model neuron. The classical surround (solid gray curve at right) is defined by the radius $r_s$ and peak amplitude $K_s$ of the larger Gaussian. The model receptive field is the linear sum of these components (dashed curve). The predicted spatial frequency tuning curve for a DOG model is obtained by convolving sinusoidal gratings of different spatial frequencies with the spatial receptive field. Alternatively, the receptive field sensitivity profile of a recorded neuron may be estimated by fitting the DOG model parameters to optimize the match to the observed spatial frequency tuning responses (SEE METHODS). (b) Each neuron's tuning curve (example cell at left) was fit to the best DOG model receptive field under three separate constraints: η<1, η=1 and η>1 (SEE METHODS). The predicted tuning curves (thin curves, right) were compared with the data, and the solution having the highest quality of fit (least $r^2$) was selected as the best model for that cell.

Because of the nature of the fitting algorithm and due to the presence of noise in the data, there is no guarantee of discovering the global minimum of the cost function. We ensured good fit by repeating the fitting process multiple (n = 100) times with random initial guesses. The quality of each of these fits was evaluated based on the Pearson correlation between the actual data and data from fits. The highest quality fit within each constraint was chosen as the candidate fit for that class of solutions. From the three candidate solutions, the final solution was taken to be the one which had the highest quality fit, except that solutions with $\eta > 1$ (surround stronger than center) were rejected if a solution with $\eta < 1, \eta = 1$ fit equally well (quality of fit within 2%). Thus we were conservative with respect to our claim that surrounds can be stronger than centers.

**Histological confirmation of recording sites**. Each subject was recorded in multiple sessions over 3-12 weeks. At the final recording session, an injection syringe loaded with 2% pontamine sky blue solution was lowered to the stereotaxic coordinates of the last recorded unit as a fiduciary mark for histological analysis. The subject was euthanized, perfused, and the brain tissue fixed, sectioned, and examined. In all cases we confirmed that the stereotaxic coordinates of the recording sites of our single units fell within the boundaries of the histologically identifiable dLGN (Discenza 2011; Paxinos G 2006).

# Results

In order to estimate the surround strength of the receptive fields we employed a standard method of fitting the spatial frequency response of a unit to a difference-of-Gaussians (DOG) model (Enroth-Cugell and Robson 1966; Grubb and Thompson 2003).

We recorded 83 well isolated single units in the putative dorso-lateral geniculate nucleus of awake head-fixed rats while they passively viewed drifting high contrast sinusoidal gratings on a linearized CRT/LCD monitor with a mean luminance of 25 cd m$^{-2}$. The monitor was approximately centered on the cell's receptive field and filled 85°X60° degrees of visual field. We varied the spatial frequency of the grating from 0.02 to 0.36 cyc/° (50 - 3 °/cyc), keeping the temporal frequency constant at 2 or 4Hz. We only include in subsequent analysis the cells whose spatial receptive field center was clearly within the confines of the monitor as measured using the spike-triggered-average to a white noise stimulus (Chichilnisky 2001), and whose responses were stationary (see Methods). Of the 83 recorded 47 fit the criteria. The recording locations were later confirmed histologically.

## Responses in awake rat dLGN to drifting gratings

Responses of one representative OFF cell are summarized in Figure 2. The raw voltage trace in response to a drifting grating for one 3-s trial (Figure 2a) shows the quality of isolation of the unit, and reveals that the firing rate was modulated over time by the stimulus (Figure 2b) at this spatial frequency. The drift speed was adjusted such that the temporal frequency of modulation was the same for all spatial frequencies. The responses to all repeats of all spatial frequencies are summarized by rasters grouped by spatial frequency (Figure 2c).

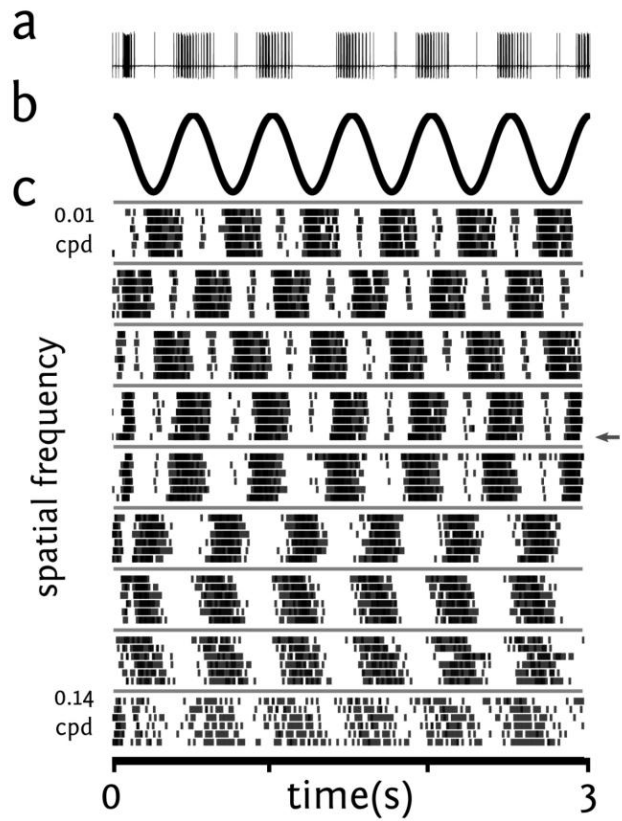

**Figure 2. Responses to drifting gratings.** (a) High-pass filtered voltage trace from one single unit recorded from the dLGN of an alert rat. The raster corresponding to this trial is marked with an arrow in (c). (b) Stimulus luminance at an arbitrary point on the display. Temporal frequency was 2 Hz, for six complete cycles within the 3 second duration of a trial, regardless of spatial frequency. (c) Rasters for this single unit obtained for all spatial frequencies, where each row indicates responses for a single trial and each tic mark indicates the time of a single action potential. The six trials recorded at each spatial frequency were interleaved during the experiment but are grouped by spatial frequency for display. The time axis at bottom applies to all panels.

For the cell shown, the mean firing rate depended on spatial frequency (Figure 3a), as was the case for N=36/47 cells in our population. The highest mean rate for any spatial frequency was 68 spikes/sec for this cell, and 22.89+/-11.27 spikes/sec (mean +/-SD) across our population. This cell showed response modulation at the temporal frequency of the stimulus (2Hz) (Figure 3b), which reflects the linear component of the response. At its optimal spatial frequency (chosen by the peak of f1), modulation of this cell was 77% of the mean rate (f1/f0). All cells in our sample were well modulated at their optimal spatial frequency: f1/f0 = 0.66+/-0.15 % (mean +/-SD across the population).

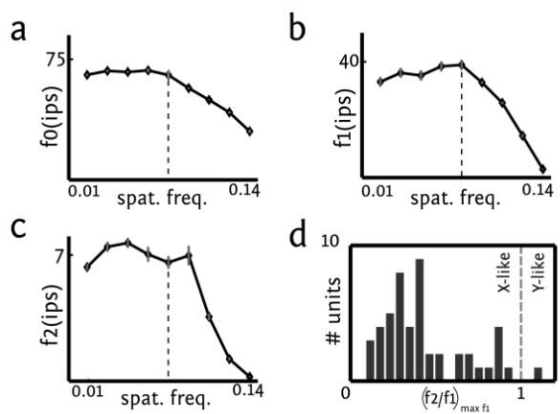

**Figure 3. Spatial frequency tuning analysis for an example cell.** (a) Average firing rate (f0, in impulses per second) as a function of spatial frequency (shown on a log scale), for the data shown in Figure 2. Diamond symbols show the mean obtained over trials, error bars (gray) show the standard error of the mean (SEM) across trials. (b) Modulation in firing rate about the mean, at the temporal frequency of the visual stimulus (f1 = 2Hz). (c) Modulation of the firing rate at twice the stimulus frequency (f2 = 4Hz). Dashed line in (a),(b) and (c) indicates the spatial frequency at which the f1 response is highest (peak in 3b). Values of f2 and f1 at this spatial frequency were used to compute the ratio f2/f1, which for this cell is 0.17. A ratio >1 indicates frequency doubling, a nonlinear characteristic of Y-like cells. (d) Distribution over the population of cells of the f2/f1 ratio measured at the spatial frequency with maximum f1 response.

This cell also had some response power at twice the input temporal frequency (f2, Figure 3c). This weak f2 response is attributable to rectification and rebound, but does not resemble a classic frequency-doubling response typical of Y cells (Hochstein and Shapley 1976a; b). Although no detailed cell classification has been attempted, we refer to this cell as "X-like" merely to indicate that at the optimal spatial frequency, f2/f1 (=0.41) was less than unity. By this definition 46/47 of the cells in our sample were X-like (Figure 3d).

**Measuring Surround Antagonism**

A standard method for measuring the extent of surround antagonism is a difference of Gaussian (DOG) model fit from spatial frequency tuning curves. We used the linear response to the grating (f1, see Figure 3b) to fit the spatial receptive field center and surround components (see Methods). For the example cell shown in Figures 2 and 3, the best fit DOG model (Figure 4a) had a center radius $r_c$ of 4°, and a surround radius $r_s$ of 13°, and a relative surround strength of η=0.83. This is typical low-pass tuning curve in our sample, and this type of response is well described in the literature. A DOG receptive field with a weak surround relative to center is consistent with strong response modulation even at the lowest spatial frequencies tested.

The measured tuning curve of another example cell is shown in Figure 4b, along with the fit obtained from the best DOG model. This is a typical band-passed tuning curve: responses fall off at both high and low spatial frequencies. This cell's receptive field model had a center radius $r_c$ of 2.19°, and a surround radius $r_s$ of 3.75°, and a relative surround strength of η=1. A DOG receptive field with a well-balanced surround relative to center is consistent with lack of response to low spatial frequencies, and is well described in the literature.

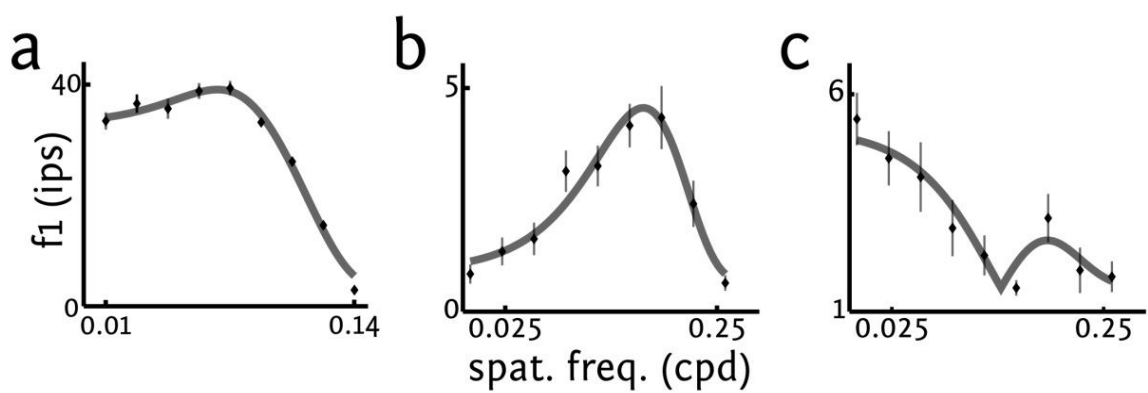

**Figure 4. Three classes of spatial frequency tuning curves in the awake rat dLGN.** In each panel, the f1 response for one recorded neuron is shown as the mean f1 over trials (diamonds) +/- SEM across trials (thin gray lines), as a function of spatial frequency (cycles per degree) on a log scale. The tuning curve predicted by the best-fit DOG model receptive field is overlaid (thick gray curve) in each case. (a) Spatial frequency tuning curve for an example low-pass tuned cell. (b) Spatial frequency tuning curve for an example band-pass tuned cell. (c) Spatial frequency tuning curve for an example notched or dual-band-pass cell.

We also found a third type of tuning curve in our sample (Figure 4c) which has not been well described previously (but see (Heine and Passaglia 2011)). These spatial frequency tuning curves could be described as dual-band-pass or notched; the notch occurred at different spatial frequencies for different cells. While DOG models with balanced (η =1) or weak (η <1) surrounds could not reproduce these tuning curves, they were easily fit by DOG models with strong surrounds (η >1). This cell's receptive field model had a center radius $r_c$ of 2.00°, and a surround radius $r_s$ of 3.97°, and a relative surround strength of η=2.2. Interpretation of these cells will be considered further in the discussion.

Most units, including the novel response type, were well fit by the classic difference-of-Gaussians (DOG) model. Quality of fit as measured by the Pearson correlation was high (>0.85) for most units (46/47). The center radius for the fits ranged between 1°-13° visual angle (5.34°+/-3.85°). This range is much broader than the range reported earlier using other methods (Fukuda et al. 1979; Hale et al. 1979). The surround radius for the fits ranged between 2.84°-125° visual angle (22.34°+/-28.8°: ), and the ratio of surround radius to center radius ranged from ~1 to ~100 (mean+/-sd: 6.43+/-15.43).

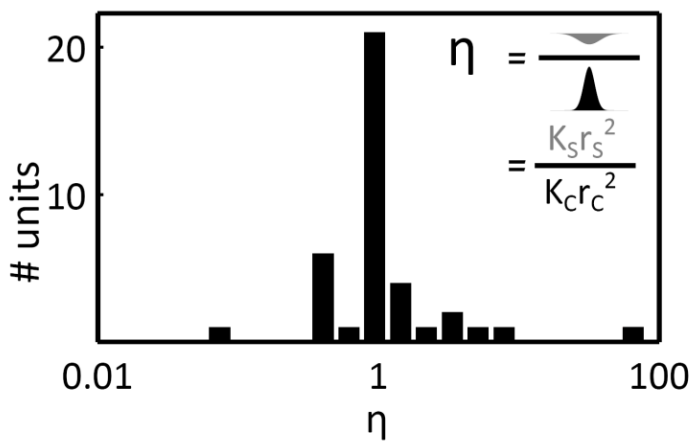

**Figure 5. Distribution of η across population.** The integrated strength of the surround relative to the center is given by η (inset equation). The distribution of η (shown on log scale) over the population is shown. The data are clustered near η=1, corresponding to well-balanced surround strength. Units with weak surrounds are at left (η<1), while those with stronger-than-balanced surrounds are at right (η>1).

We measured the integrated weight of the surround relative to the center by the variable η (Equation 1, Methods). In our data values of η ranged from η=0.2 to η=80 (Figure 5). The value of η corresponded closely to the shape of the spatial frequency tuning curve. All cells that showed significant fall-off of the response at low spatial frequencies had η values close to 1 (Figure 4b). Those that showed little fall-off at low spatial frequencies had η values less than 1 (Figure 4a), and all cells that showed notched spatial frequency tuning curves had η values greater than 1 (Figure 4c).

The majority of cells (N=29/47) had well balanced center and surround with 0.95 <= η <= 1 05.  Several (N=8/47) had weak surrounds (η < 0.95) as previously described in other studies. A substantial fraction of cells (10/47) had values η > 1.05, indicating a surround that is stronger than the center. Because this result was unexpected, our analysis was conservative in assigning fits with η > 1: we searched separately for the best fit model with η = 1 and with η < 1, and chose one of these solutions preferentially if the fit was almost as good (see Methods). We emphasize that this reflects the *integrated* weight of the center and surround over space, and the surround radius was often much larger. The peak *sensitivity* of the surround $K_s$ was less than that of the center $K_c$ in all cases.

## Discussion

We have used spatial frequency tuning to estimate the spatial structure of receptive fields in the dLGN of un-anesthetized rats. The distribution of receptive field center sizes we find is in good agreement with that previously reported for retinal ganglion cells of the rat (Heine and Passaglia 2011). Some neurons in our sample had weak surrounds, as reported in other species. We find that most dLGN neurons in the alert rat, however, have well-balanced surround antagonism and thus a band-pass rather than low-pass spatial frequency tuning. This differs from the distributions reported in other species, in which surrounds are typically weaker (Cheng et al. 1995; O'Keefe et al. 1998; Xu et al. 2002). The distribution of surround strength has not previously been reported in either anesthetized or alert rodents. In anesthetized mouse dLGN, both low-pass and band-pass spatial frequency tuning curves have been observed (Grubb and Thompson 2003) consistent with receptive fields ranging from weak to balanced. In optic nerve recordings in anesthetized rats, retinal ganglion cell spatial frequency tuning curves were found to be mostly low-pass, although band-pass tuning curves were also observed (Heine and Passaglia 2011).

We also report several examples of cells with unexpected, notched or dual-band-pass tuning curves which have not been reported in other species (but see Figure 7 in (Heine and Passaglia 2011)). These data could be well fit by the standard circularly concentric DOG model if the integral of the surround component exceeded that of the center (Figure 4c). For an intuition, consider a set of model cells with center size 1°, surround size 3°, and different surround strengths. Figure 6 shows the response of such model cells to drifting gratings, separated into center and surround components. The center response (C) and the surround responses (S) are both decreasing functions of spatial frequency. As the surround strength increases, the response at the lowest spatial frequency for center and surround get closer in value. The net response at lower frequencies (C-S) diminishes. For the case where the surround is stronger than the center, the net response (C-S) reaches zero at an intermediate spatial frequency and takes negative values, corresponding to a phase reversal in the grating response. However, the value that is experimentally measured is the amplitude of the response modulation about the mean, which is always a positive value. Therefore when center and surround responses are summed, a notched or dual-band-pass tuning curve is produced. We take this to be a simple and conservative explanation of the data, though other models are possible, such as an antagonistic surround that is spatially offset relative to the receptive field center (Soodak 1986). Regardless of how these tuning curves arise mechanistically, such cells clearly do occur in our data.

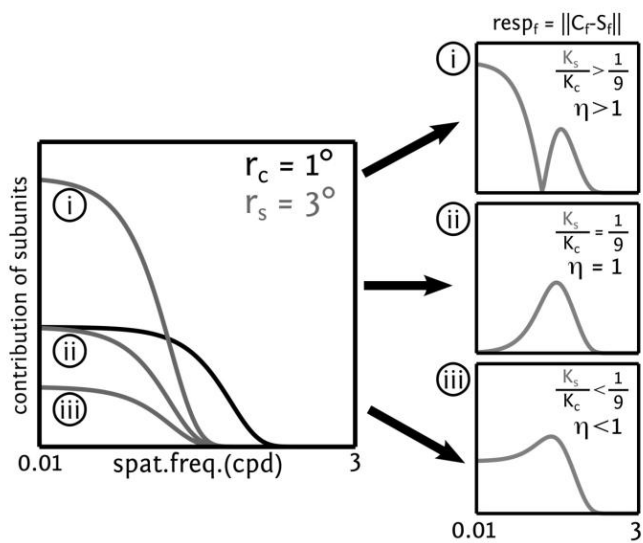

**Figure 6. Intuition for notched or dual-band-pass tuning curves.** Responses of the center and surround components to drifting gratings of different spatial frequencies for three hypothetical cells (left). Center response (black curve) is for a Gaussian profile with radius $r_c$ of 1° and arbitrary sensitivity of 1, and is the same for all three model cells. Surround responses (grey curves) are for Gaussian profiles with $r_s$ = 3° and different sensitivities relative to the center. When the integrated surround strength is less than that of the center (iii), the response reaches a plateu at low spatial frequency. When the center and surrounds balance (ii), response falls to 0 at low spatial frequency. If the surround response exceeds that of the center (i), a node in the tuning curve results.

**Linearity of Responses**

Difference-of-Gaussian models based on spatial frequency tuning data are widely used to describe spatial receptive field structure of both X and Y cells in the retina and dLGN (Bonin et al. 2005; Linsenmeier et al. 1982; Sceniak et al. 2006). If the combined circuitry leading from the visual image to a dLGN neuron's response were a perfectly linear system, response due to the center and surround components to an arbitrary pattern of light would be separable and add linearly, and this method of measurement would be exact. To the extent that responses in the dLGN are nonlinear, this method provides only an approximation or a description of the linear component of the response. Not surprisingly, past studies report that better fits are obtained for X cells than for Y cells (Linsenmeier et al. 1982).

For the cells in our study we do not have data from standing phase reversing gratings or sparse noise, which would be required to determine whether the receptive fields contained nonlinear spatial subunits. Based on the ratio of the f2 to f1 responses to drifting gratings, most of the neurons in our population had relatively linear responses. In this specific sense, we refer to our cells as "X-like" (Figure 3d), and we find our data to be well fit by the DOG model. We note that additional classes of neurons in the dLGN with other response properties, including nonlinear or Y-like cells, might exist but could be missed due to an unknown selection bias in our recording technique.

Another known form of nonlinearity in the dLGN is extra-classical surround suppression, which is modeled as a divisive normalization component (Bonin et al. 2005; Heeger 1992) . Our drifting grating stimuli were large compared to the classical receptive field centers, and therefore may have engaged extraclassical surround suppression. Indeed we find that responses in the dLGN can be sensitive to the size of the stimulus aperture (our unpublished data). Divisive normalization, if present, would affect the absolute amplitude ($K_c$, $K_s$) of our DOG fit but should not impact our measures of center or surround radius ($r_c$, $r_s$) nor the relative sensitivity ($K_s/K_c$) or surround strength (η).

**Location of computation**

We have measured center-surround antagonism in the dLGN of the alert rat. This property is at least partly inherited from the retinal inputs to the dLGN. We do not have access to a comparable measurement of surround strength in rat retinal ganglion cells, so it remains to be determined if thalamic or cortico-thalamic circuitry contribute to surround strength in the dLGN of the alert rat. In other preparations, however, surrounds of dLGN receptive fields are thought to be stronger than that of their retinal inputs (Cheng et al. 1995; Hubel and Wiesel 1961). Anesthesia is known to affect dLGN responses; we cannot exclude the possibility that anesthesia affects surround strengths in the dLGN.

**Computational Function**

Our results imply that in the alert rat, most cells in the dLGN are transmitting a signal that is spatially decorrelated, and therefore less redundant and more sparse than the luminance patterns found in natural scene scenes (Barlow 1961; Olshausen and Field 1996). It remains unclear whether the resulting compression of the image is the most important function of this spatial filtering operation, given that the combined effect of eye movements already tends to spatially decorrelate or "whiten" the image prior to encoding by photoreceptors, at least in some species (Kuang et al. 2012; Reinagel and Zador 1999). Alternatively, the function may be primarily the enhancement of edges: well-balanced surround antagonism would maximally enhance the difference between inputs to ON and OFF subregions of V1 receptive fields along edges and contours, serving to facilitate edge detection.

A smaller population of cells had notched or dual-band-pass spatial filters, which can be explained by extra-strong surrounds in the classical center-surround receptive field. This property may be inherited from retinal ganglion cells (Heine and Passaglia 2011). It remains to be determined whether these unusual tuning curves are a functional adaptation to visual coding, or merely a consequence of imperfect wiring. In either case, this response type comprises a substantial fraction of our sampled population, so it will be important to determine their impact on the neural code of the rat dLGN.

Another subpopulation of cells in our sample had weak surround antagonism, as widely reported in other species. These cells carry a smoothed image representation that would be spatially correlated for natural images, with little or no edge enhancement. Perfect surround antagonism removes all information about absolute luminance from the image representation; a small subset of neurons with weak or absent surrounds would be sufficient to carry this complementary information.

We find heterogeneity of surround architecture in the dLGN population, which may be functionally important (Soo et al. 2011). Past theories derived the optimal surround strength for a homogenous population of units under different stimulus conditions; it would be interesting to extend this approach to consider optimal distributions in a heterogeneous population.

## Acknowledgements

The authors thank Sarah Petruno for expert technical assistance; David Kleinfeld's laboratory for providing electrodes and technical advice; Ed Callaway for discussion and advice in design of the study; and other members of the Reinagel laboratory, who participated in developing the hardware and software used in the collection of these data. This work was supported by the National Institutes of Health (R01-EY016856-02), the San Diego Foundation (Blasker Award), and the James S. McDonnell Foundation (Scholar Award).

# Legends

**Figure 1. Difference of Gaussian Model Fitting.** (a) A model receptive field is defined by two circular, concentric 2D Gaussian densities, a cross section through which is shown at right. The receptive field center (solid black curve) is defined by the radius $r_c$ and peak amplitude $K_c$ of the smaller Gaussian, the sign of which determines the response type (ON or OFF) of the model neuron. The classical surround (solid gray curve at right) is defined by the radius $r_s$ and peak amplitude $K_s$ of the larger Gaussian. The model receptive field is the linear sum of these components (dashed curve). The predicted spatial frequency tuning curve for a DOG model is obtained by convolving sinusoidal gratings of different spatial frequencies with the spatial receptive field. Alternatively, the receptive field sensitivity profile of a recorded neuron may be estimated by fitting the DOG model parameters to optimize the match to the observed spatial frequency tuning responses (SEE METHODS). (b) Each neuron's tuning curve (example cell at left) was fit to the best DOG model receptive field under three separate constraints: $\eta<1$, $\eta=1$ and $\eta>1$ (SEE METHODS). The predicted tuning curves (thin curves, right) were compared with the data, and the solution having the highest quality of fit (least $r^2$) was selected as the best model for that cell.

**Figure 2. Responses to drifting gratings.** (a) High-pass filtered voltage trace from one single unit recorded from the dLGN of an alert rat. The raster corresponding to this trial is marked with an arrow in (c). (b) Stimulus luminance at an arbitrary point on the display. Temporal frequency was 2 Hz, for six complete cycles within the 3 second duration of a trial, regardless of spatial frequency. (c) Rasters for this single unit obtained for all spatial

frequencies, where each row indicates responses for a single trial and each tic mark indicates the time of a single action potential. The six trials recorded at each spatial frequency were interleaved during the experiment but are grouped by spatial frequency for display. The time axis at bottom applies to all panels.

**Figure 3. Spatial frequency tuning analysis for an example cell.** (a) Average firing rate ($f0$, in impulses per second) as a function of spatial frequency (shown on a log scale), for the data shown in Figure 2. Diamond symbols show the mean obtained over trials, error bars (gray) show the standard error of the mean (SEM) across trials. (b) Modulation in firing rate about the mean, at the temporal frequency of the visual stimulus ($f1 = 2Hz$). (c) Modulation of the firing rate at twice the stimulus frequency ($f2 = 4Hz$). Dashed line in (a),(b) and (c) indicates the spatial frequency at which the f1 response is highest (peak in 3b). Values of f2 and f1 at this spatial frequency were used to compute the ratio $f2/f1$, which for this cell is 0.17. A ratio >1 indicates frequency doubling, a nonlinear characteristic of Y-like cells. (d) Distribution over the population of cells of the f2/f1 ratio measured at the spatial frequency with maximum f1 response.

**Figure 4. Three classes of spatial frequency tuning curves in the awake rat dLGN.** In each panel, the f1 response for one recorded neuron is shown as the mean f1 over trials (diamonds) +/- SEM across trials (thin gray lines), as a function of spatial frequency (cycles per degree) on a log scale. The tuning curve predicted by the best-fit DOG model receptive field is overlaid (thick gray curve) in each case. (a) Spatial frequency tuning curve for an example low-

pass tuned cell. (b) Spatial frequency tuning curve for an example band-pass tuned cell. (c) Spatial frequency tuning curve for an example notched or dual-band-pass cell.

**Figure 5. Distribution of η across population**. The integrated strength of the surround relative to the center is given by η (inset equation). The distribution of η (shown on log scale) over the population is shown. The data are clustered near η=1, corresponding to well-balanced surround strength. Units with weak surrounds are at left (η<1), while those with stronger-than-balanced surrounds are at right (η>1).

**Figure 6. Intuition for notched or dual-band-pass tuning curves.** Responses of the center and surround components to drifting gratings of different spatial frequencies for three hypothetical cells (left). Center response (black curve) is for a Gaussian profile with radius $r_c$ of 1° and arbitrary sensitivity of 1, and is the same for all three model cells. Surround responses (grey curves) are for Gaussian profiles with $r_s$ = 3° and different sensitivities relative to the center. When the integrated surround strength is less than that of the center (iii), the response reaches a plateau at low spatial frequency. When the center and surrounds balance (ii), response falls to 0 at low spatial frequency. If the surround response exceeds that of the center (i), a node in the tuning curve results.